\documentclass[aip,jap,twocolumn,supercriptaddress,10pt]{revtex4-1}
\usepackage{amssymb}
\usepackage{amsfonts}
\usepackage{amsmath}
\usepackage{color}

\usepackage[pdftex]{graphicx}

\begin{document}

\date{\today}

\title{Microwave heating-induced DC magnetic flux penetration in YBa$_{2}$Cu$_{3}$O$_{7-\delta}$ superconducting thin films}

\author{Julien Kermorvant}
\affiliation{Unit\'{e} Mixte de Recherche en Physique, CNRS UMR 137, THALES Research and Technology, 1 Boulevard Augustin Fresnel, 91125 Palaiseau cedex, France}

\author{Jean-Claude Mage}
\affiliation{Unit\'{e} Mixte de Recherche en Physique, CNRS UMR 137, THALES Research and Technology, 1 Boulevard Augustin Fresnel, 91125 Palaiseau cedex, France}

\author{Bruno Marcilhac}
\affiliation{Unit\'{e} Mixte de Recherche en Physique, CNRS UMR 137, THALES Research and Technology, 1 Boulevard Augustin Fresnel, 91125 Palaiseau cedex, France}

\author{Yves Lema\^{i}tre}
\affiliation{Unit\'{e} Mixte de Recherche en Physique, CNRS UMR 137, THALES Research and Technology, 1 Boulevard Augustin Fresnel, 91125 Palaiseau cedex, France}

\author{Jean-Fran\c{c}ois Bobo}
\affiliation{CEMES, CNRS UPR 8011 \& Universit\'{e} de Toulouse, 29 rue Jeanne Marvig, 31055 Toulouse, France}
\affiliation{ONERA DEMR, 2 avenue Edouard Belin, 31055 Toulouse cedex 4, France}

\author{Cornelis Jacominus van der Beek}
\affiliation{Laboratoire des Solides Irradi\'{e}s, CNRS-UMR 7642 \& CEA-DSM-IRAMIS, Ecole Polytechnique, F 91128 Palaiseau cedex, France}

\date{\today}

\begin{abstract}
The magneto-optical imaging technique is used to visualize the penetration of the magnetic induction in YBa$_{2}$Cu$_{3}$O$_{7-\delta}$  thin films during surface resistance measurements. The in-situ surface resistance measurements were performed at 7 GHz using the dielectric resonator method. When only the microwave magnetic field $H_{rf}$ is applied to the superconductor, no $H_{rf}$-induced vortex penetration is observed, even at high rf power. In contrast, in the presence of a constant magnetic field superimposed on $H_{rf}$ we observe a progression of the flux front as $H_{rf}$ is increased. A local thermometry method based on the measurement of the resonant frequency of the dielectric resonator placed on the YBa$_{2}$Cu$_{3}$O$_{7-\delta}$  thin film shows that the $H_{rf}$--induced flux penetration is due to the increase of the film temperature.
\end{abstract}

\maketitle

  \section{ Introduction}

High Temperature Superconductor (HTS) thin films are now recognized as particularly suitable for high frequency signal processing. Due to their very low surface resistance $R_{s}$, as compared to normal metals, they allow for very efficient microwave signal filtering and detection. However, in the region of high power $P_{rf}$ of the incident microwave field, their application is limited by the strong dependence of $R_{s}$ on the $P_{rf}$--magnitude. A nonlinear increase of the surface resistance with the input rf power is commonly observed,\cite{Oates92,Oates95,Samoilova95,Hampel1996,Wosik97,Anlage99,Lahl2005,Kermorvant2009} leading to a detrimental decrease of the $Q$-factor of the devices.\cite{Hein97} The origin of the nonlinear microwave losses in high $T_{c}$ superconductors has been studied by many groups. Among the cited causes, there are intrinsic phenomena such as the excitation of quasiparticles when the  current density induced by the microwave magnetic field (of magnitude $H_{rf}$) becomes of the order of the pair-breaking current density,\cite{Dahm99} but also the limitation of the supercurrent in grain boundaries, and vortex motion \cite{Lahl2005,Coffey1992,Coffey1993,Coffey1993ii} in the superconductor induced by the rf field. The interplay of the superposed ac and dc magnetic fields in superconducting thin films is extensively described in Refs. \onlinecite{Brandt2002,Mikitik2003,Mikitik2004,Mikitik2005}. Recent work has, however, clearly demonstrated that it is simply local Joule heating of the superconducting film by the microwave field that leads to the nonlinear behavior,\cite{Hampel1996,Wosik97,Kermorvant2009} and that the dissipation at the origin of the heating is due to the linear electromagnetic response of the films. 

In Ref.~\onlinecite{Kermorvant2009}, we have introduced a valuable tool for  local thermometry of superconducting films studied using a dielectric resonator. The variation of the resonator frequency as a function of the temperature depends essentially on the dielectric constant $\epsilon$ of the resonator; its calibration turns the latter into a precise local thermometer. This method has allowed us to measure the temperature of YBa$_{2}$Cu$_{3}$O$_{7-\delta}$ films as a function of the rf input power (at 10 GHz) under nominally isothermal conditions. The observed temperature increase of the resonator and the YBa$_{2}$Cu$_{3}$O$_{7-\delta}$ thin films as function of rf power unambiguously showed  that  the usually observed increase in the surface resistance is due to Joule heating, with a linear response dissipation mechanism. Candidate mechanisms at the origin of the heating can therefore be limited to quasi-particle dissipation \cite{Dahm99,Mattis1958} and flux-flow losses.\cite{Coffey1992,Coffey1993,Coffey1993ii,Bardeen1965,Brandt1991,Coffey1991}

 \begin{figure}[b]%
\includegraphics[width=1.01\linewidth]{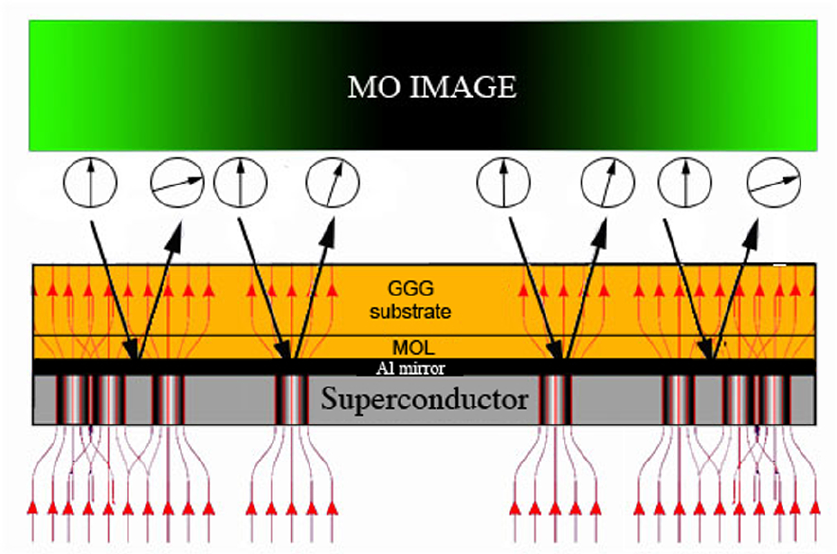}%
\caption{(Color online) Principle of MOI of superconductors. Thin drawn lines depict the magnetic flux as this traverses the superconductor and the MOL, thick black lines show the optical path of the impinging and reflected light, and the circled arrows illustrate the linear polarization direction of the light.}
\label{fig2}%
\end{figure}
 
 \begin{figure}[t]%
\vspace{-0mm}
\includegraphics[width=1.01\linewidth]{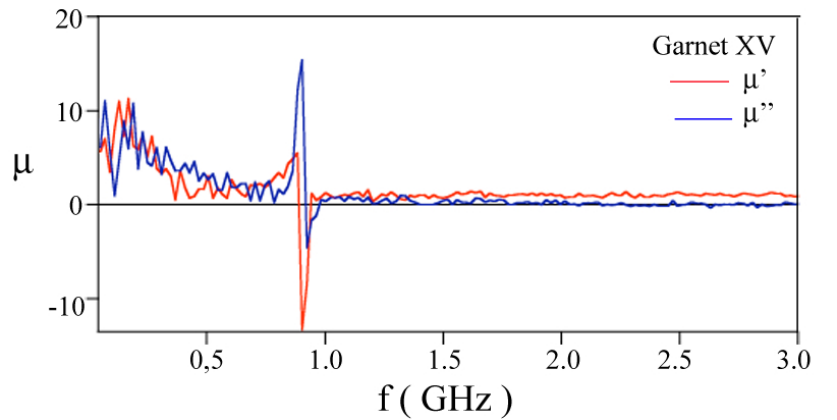}%
\vspace{-0mm}
\caption{Complex permeability of the (Lu,Bi)$_{3}$(Fe,Ga)$_{5}$O$_{12}$  magneto-optically active layer, measured at room temperature.}
\label{fig1}%
\end{figure}

Here, we study the influence of an rf magnetic field generated by a rutile dielectric resonator on the DC flux distribution in YBa$_{2}$Cu$_{3}$O$_{7-\delta}$ thin films. For this, we have developed a dedicated  set-up, that allows for the simultaneous measurement of the film surface resistance and the visualization of the magnetic flux distribution using the Magneto-Optical Imaging (MOI) Technique.\cite{Dorosinskii1992,Jooss2002} MOI is based on the Faraday effect in which the rotation of the polarization plane of  incident linearly polarized light is proportional to the magnetic induction component parallel to the wavevector of the incoming light. Since superconductors do not present a significant Faraday effect, one has to use a magneto-optical layer (MOL, with a strong Faraday effect) placed on top of a superconductor, as depicted in Fig.~\ref{fig2}. In the experiments described below, we use, as a MOL, Lu- and Bi-doped Yttrium-Iron Garnet thick films\cite{Uehara2009} with a ferromagnetic resonance at 0.9 GHz (see Fig.~\ref{fig1}).  The application of a magnetic field of a few dozen mT will increase the ferromagnetic resonance frequency somewhat. However, at the experimental microwave field frequency of 7.0 GHz, the magnetization rotation leading to the Faraday effect will always be strongly overdamped, inhibiting the direct visualization of the microwave magnetic field. The MOI technique does permit the visualization of the modification of static flux structures in YBa$_{2}$Cu$_{3}$O$_{7-\delta}$ thin films following microwave field application. Note that the development of a similar imaging system was reported  in Ref.~\onlinecite{Wosik2001}, however, the influence of the microwave magnetic field on flux penetration was not reported there.

 \begin{figure}[t]%
\includegraphics[width=0.9\linewidth]{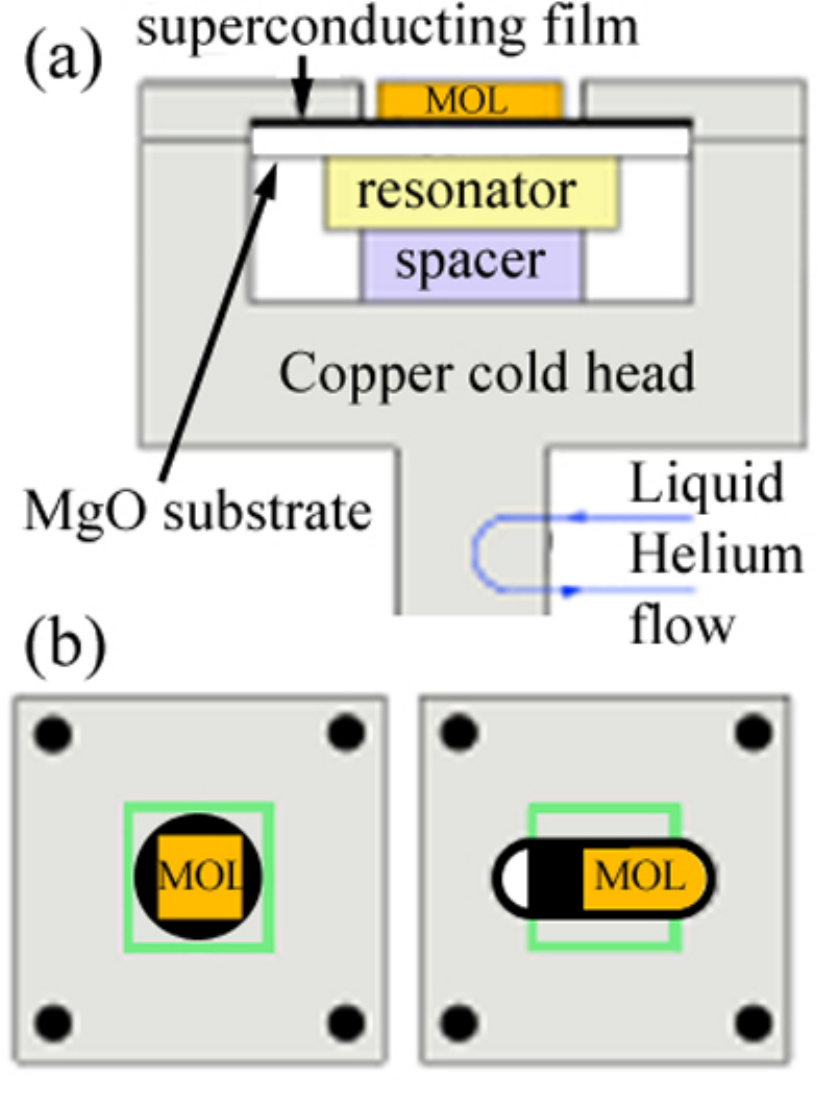}%
\caption{(Color online) Schematic view of the experimental assembly. Panel (a) shows a side view, while panel (b) depicts a top view for different lid apertures.  The left-hand side cover allows the imaging of the central region of the sample, while the right-hand cover allows for the observation of the edge region. }
\label{fig3}%
\end{figure}

\section{Experimental details}
 
\subsection{YBa$_{2}$Cu$_{3}$O$_{7-\delta}$ films }

All YBa$_{2}$Cu$_{3}$O$_{7-\delta}$ films under study in this work were cut from the same wafer, designated SY211 in Ref.~\onlinecite{Kermorvant2009}. The YBa$_{2}$Cu$_{3}$O$_{7-\delta}$ was deposited on a 500 $\mu$m-thick MgO substrate using cylindrical hollow cathode DC sputtering, and had a thickness $d =400$ nm. Its critical temperature is $T_{c} \simeq 86$ K and its critical current density $j_{c} = 5\times 10^{10}$ Am$^{-2}$ at $T = 77$ K.

\subsection{Magneto-Optical Imaging}

The MOL is a 5 $\mu$m-thick (Lu,Bi)$_{3}$(Fe,Ga)$_{5}$O$_{12}$ ferrimagnetic garnet film with in-plane magnetic anisotropy, grown on a 500 $\mu$m-thick Gd$_{3}$Gd$_{5}$O$_{12}$ (GGG) substrate.\cite{Uehara2009} The MOL is covered by a 100 nm-thick Al mirror layer and a 10 nm-thick TiO$_{2}$ protective layer. It is placed face-down on the superconducting film (Fig.~\ref{fig2}), whence it is observed through the transparent substrate using a polarized microscope with nearly crossed polarizers. In this configuration, the reflected light intensity increases as function of the local magnetic induction perpendicular to the garnet. Bright regions in the MO image correspond to regions of high magnetic flux density, while dark areas correspond to small or zero induction. This allows for the direct observation of magnetic flux penetration into the YBa$_{2}$Cu$_{3}$O$_{7-\delta}$ films. The calibration of the luminous intensity in the absence of the superconductor, or measured at a point that is sufficiently far removed from the superconductor, allows one to convert the spatially resolved intensity maps to maps of  the absolute value of the magnetic induction. 

\subsection{Simultaneous MOI and Surface Resistance measurements}

In order to measure the YBa$_{2}$Cu$_{3}$O$_{7-\delta}$ film's surface resistance during  MOI, a number of specific modifications are required with respect to standard MOI and standard $R_{s}$ measurements. These are sketched in Fig.~\ref{fig3}. First of all, we have chosen to use the dielectric resonator technique with a rutile-phase TiO$_{2}$ resonator (of diameter 7 mm and height 3 mm) operating at 10 GHz for the $R_{s}$ measurements.\cite{Kermorvant2009}  However, the presence of the MOL, in close contact with the superconducting film, prohibits one from placing the dielectric resonator directly on the film. The resonator was therefore installed on the side of the MgO substrate. The resonator is pressed to the substrate by the bottom lid of the square Cu cavity (of width 30 mm and height 7 mm) in which the whole assembly is placed.  The presence of the substrate between the resonator and the film shifts the resonant frequency from 10 GHz to 7 GHz. The MgO substrate exhibits a low microwave loss tangent, $\tan \delta = 9 \times 10^{-6}$, and high thermal conductivity, $\kappa(90\,\,{\mathrm  K}) = 290$ Wm$^{-1}$K$^{-1}$.\cite{Cahill} Hence, it does not introduce additional losses, and negligible measurement error on both the temperature and the resonant frequency when the temperature of the superconductor rises. The microwave field is excited using the sweeper of a HP 8510C vector network analyzer through a coupling loop in the Cu cavity, the position of which can be adjusted at low temperature using an XYZ stage.

\begin{figure}[b]%
\includegraphics[width=0.99\linewidth]{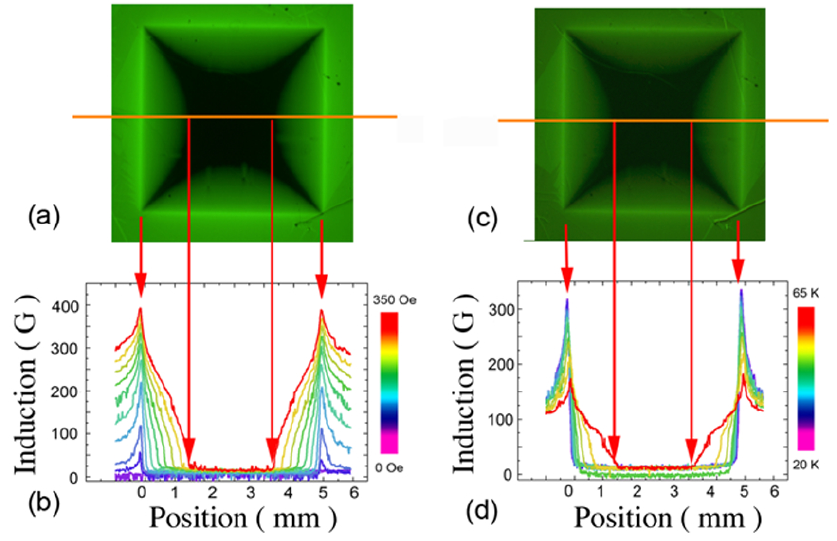}
\caption{(Color online) (a) MOI of the flux distribution in the square YBa$_{2}$Cu$_{3}$O$_{7-\delta}$ film at $H_{a} = 350$ Oe and $T = 50$ K. (b) Profiles of the magnetic induction $B$ along the horizontal line in (a), for successive applied fields between 0 and 250 Oe.  (c) MOI of the flux distribution in the film at $T = 70$ K and $H_{a} =150$ Oe, applied after zero field cooling to 30 K and successive warming. (d) $B$--profiles taken along the horizontal line in (c),  for different temperatures during warming to 70 K. }
\label{fig4}%
\end{figure}

In usual MOI, the object to be imaged is cooled via direct thermal contact with a Cu sample holder attached to the cold head of the cryostat. For simultaneous $R_{s}$/MOI measurements, the intercalation of the dielectric resonator, with its low thermal conductivity,\cite{Thurber1965} prohibits this configuration. Thus, thermal contact is made via the top of the imaged specimen, through the top lid of the Cu cavity. This imposes a reduction of the magneto-optically imaged area of the sample under study. Imaging is performed through a circular  aperture of diameter 8 mm in the cavity top lid.  The presence of the Al mirror layer on the MOL does not pose additional problems, since it is shielded from the TiO$_{2}$ resonator by the YBa$_{2}$Cu$_{3}$O$_{7-\delta}$ film.

\section{Results}

\subsection{ MOI in dc magnetic field}

Figures~\ref{fig4}(a,c) present standard MO images obtained on a square-shaped $5\times5$ mm$^{2}$ YBa$_{2}$Cu$_{3}$O$_{7-\delta}$ film. The panels (a,c) illustrate the distribution of the magnetic induction in the superconducting film during the application of a constant magnetic field  $H_{a} = 350$ Oe perpendicular to the film plane at the cavity temperature of 50 K, while panels (b,d) show the effect of increasing the cavity temperature after zero field cooling to 30 K, the application of $H_{a} = 150$ Oe and subsequent warming to 70 K.  In all cases, the flux distribution accurately corresponds to the predictions of the Bean model,\cite{Bean1962,Brandt1993,Zeldov1994} which has that in the flux-penetrated areas of the film, the screening current density can only take on the value $\pm  j_{c}$. The magnetic flux penetrates the superconductor from the edge of the sample, and is distributed according to the characteristic pillow-like shape expected for the divergence-free flow of the critical current in the thin film.\cite{Brandt1995} The flux distribution only depends on the parameter $H_{a}/j_{c}$, so that increasing either $H_{a}$ or temperature (with the concomitant decrease of $j_{c}$) both lead to a progression of the flux front to the film center. The position of the flux front, $x_{f}$, verifies the relation 
\begin{equation}
x_{f} = w \left[ 1 - \frac{1}{\cosh \left( \pi H_{a} /j_{c} d \right)} \right]
\end{equation}
in which $w$ is the half-width of the YBa$_{2}$Cu$_{3}$O$_{7-\delta}$ film.

\begin{figure}[t]%
\includegraphics[width=0.9\linewidth]{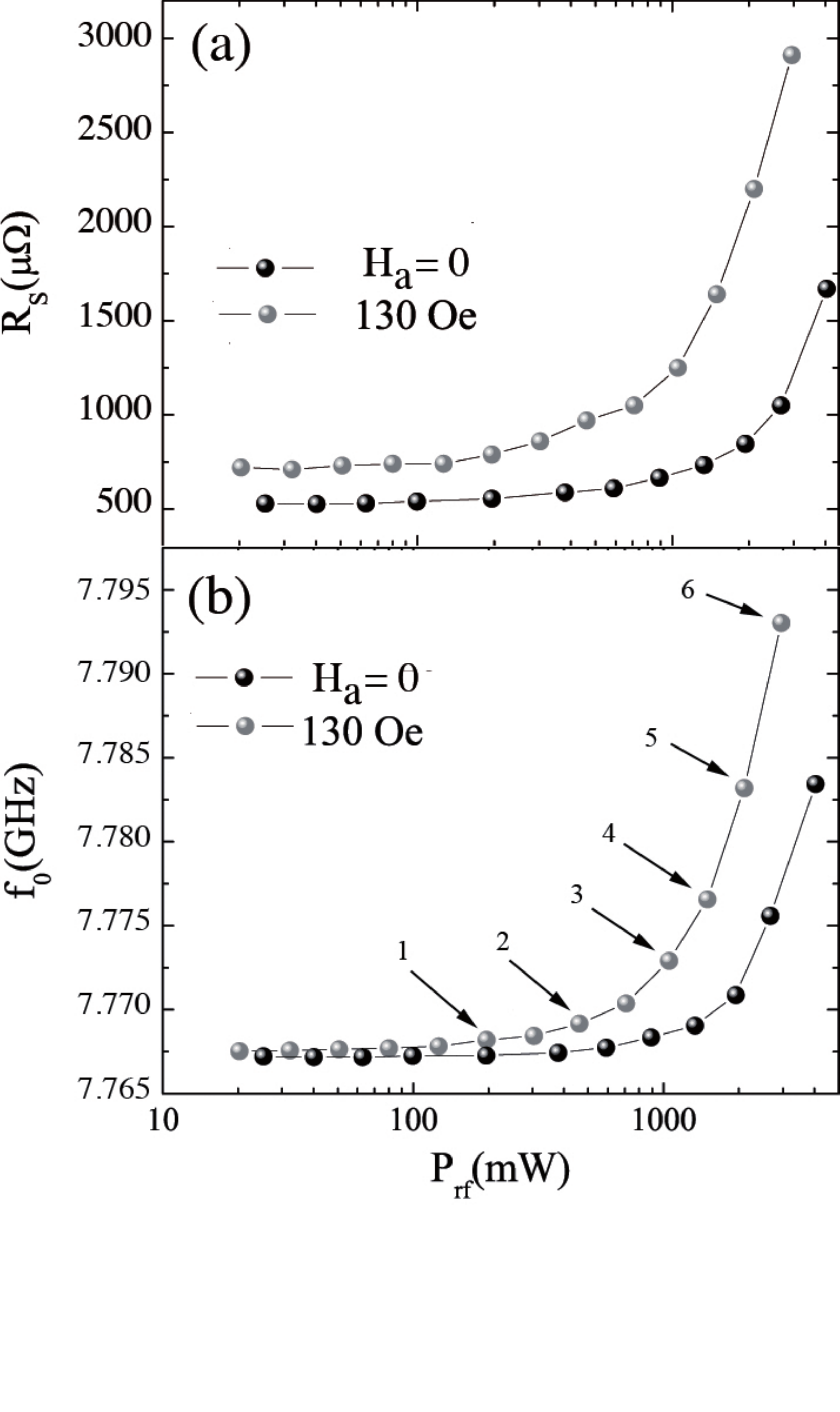}%
\vspace{-6mm}
\caption{ (a) Surface resistance at 7 GHz as function of the input microwave power at 60 K, for zero applied field, and $H_{a} = 130$ Oe. (b) Resonant frequency as a function of the input microwave power at 60 K. }
\label{fig5}%
\end{figure}

\begin{figure}[t]%
\includegraphics[width=1.0\linewidth]{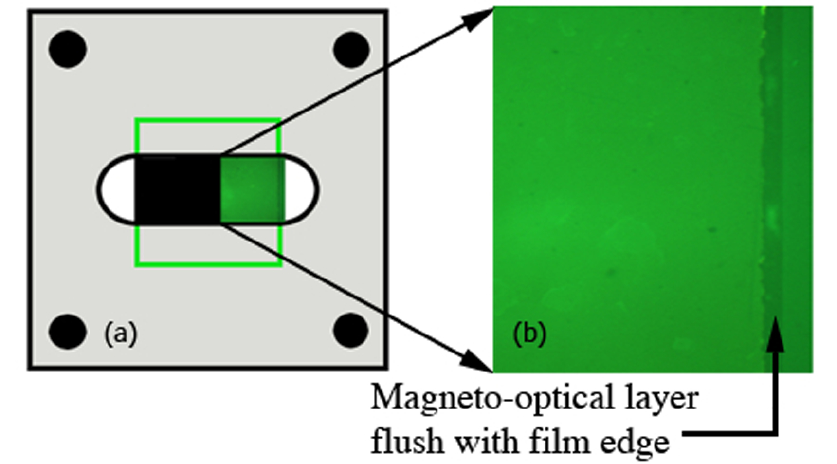}%
\caption{(Color online) (a) Schematic representation of the position of the obtained image relatively to the sample edge. The open green contour denotes the outline of the SC film. (b) MO image obtained after the sample was zero field cooled to 60 K, and a microwave field with input power of 22 dBm was applied. }
\label{fig6}%
\end{figure}

\begin{figure}[b]%
\includegraphics[width=1.0\linewidth]{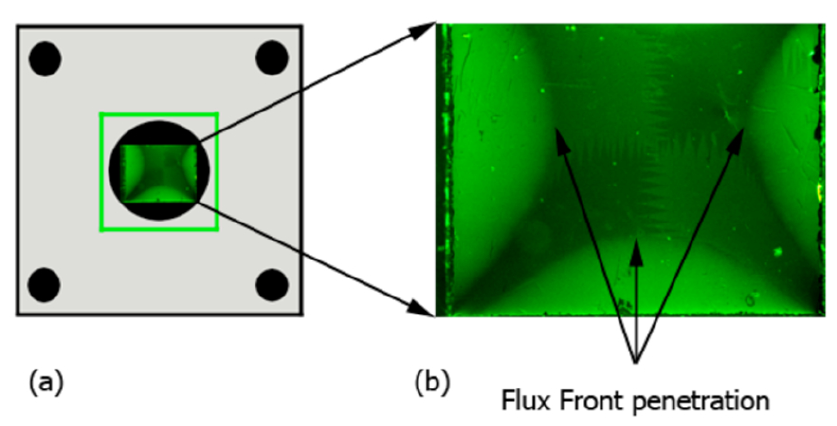}

\caption{(Color online) (a) Schematic representation of the position of the obtained image relative to the YBa$_{2}$Cu$_{3}$O$_{7-\delta}$  thin film outline (open green contour). (b) MO image obtained after the sample was zero-field cooled to 60 K, and a dc field of 130 Oe applied.}
\label{fig7}%
\end{figure}

\subsection{MOI with applied microwave magnetic field}

In a first experiment, we have applied a microwave field to the superconductor, in the absence of a dc field. Since any magnetic flux should penetrate the superconducting film from the edge, we have used the right-hand Cu cavity cover  of Fig.~\ref{fig3}.  The film was cooled in Earth's magnetic field, down to 60 K. After stabilization of the target temperature, the $TE_{011}$--mode of the dielectric resonator was excited in the low-power regime (-10 dBm applied to the coupling loop). In this mode, the rf electric field is parallel to the plane of the superconducting film, while the rf magnetic field is perpendicular to it. Next, the microwave power was increased from -10 dBm to +24 dBm in steps of 3 dBm. At each input power level, the resonator frequency and $Q$-factor were measured and the corresponding MO image recorded.

Figure~\ref{fig5}(a) renders the microwave power dependence of the surface resistance $R_{s}$, while Fig.~\ref{fig5}(b) shows  the power dependence of the resonant frequency $f_{0}$. No qualitative difference was observed with respect to the previously performed measurements at 10 GHz, with the resonator placed in direct contact with the superconducting film.\cite{Kermorvant2009} Both $R_{s}$ and the resonant frequency increase with rf power, indicating the nonlinear increase of microwave losses in the high microwave power regime. 

Figure~\ref{fig6}(b) shows the MO image obtained at an input power level of 24 dBm. According to the $R_{s}$($P_{rf}$) characteristics in Fig.~\ref{fig5}(a), the superconductor is clearly in the lossy regime. However, MOI does not reveal any vortex penetration into the superconductor, even in the high power regime. 

 \begin{figure}[t]%
\includegraphics[width=0.9\linewidth]{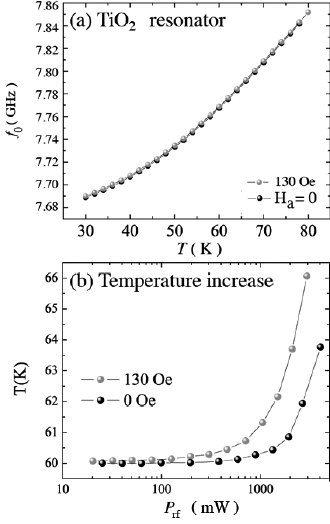}
\caption{(a) Increase of the resonance frequency of the TiO$_{2}$ resonator as function of temperature, measured in the absence of the superconducting film. (b) Temperature increase of the superconducting film during the swept-power experiments of Fig.~\protect\ref{fig5}, as deduced from (a). }
\label{fig8}%
\end{figure}

 \begin{figure}[t]%
\includegraphics[width=1.0\linewidth]{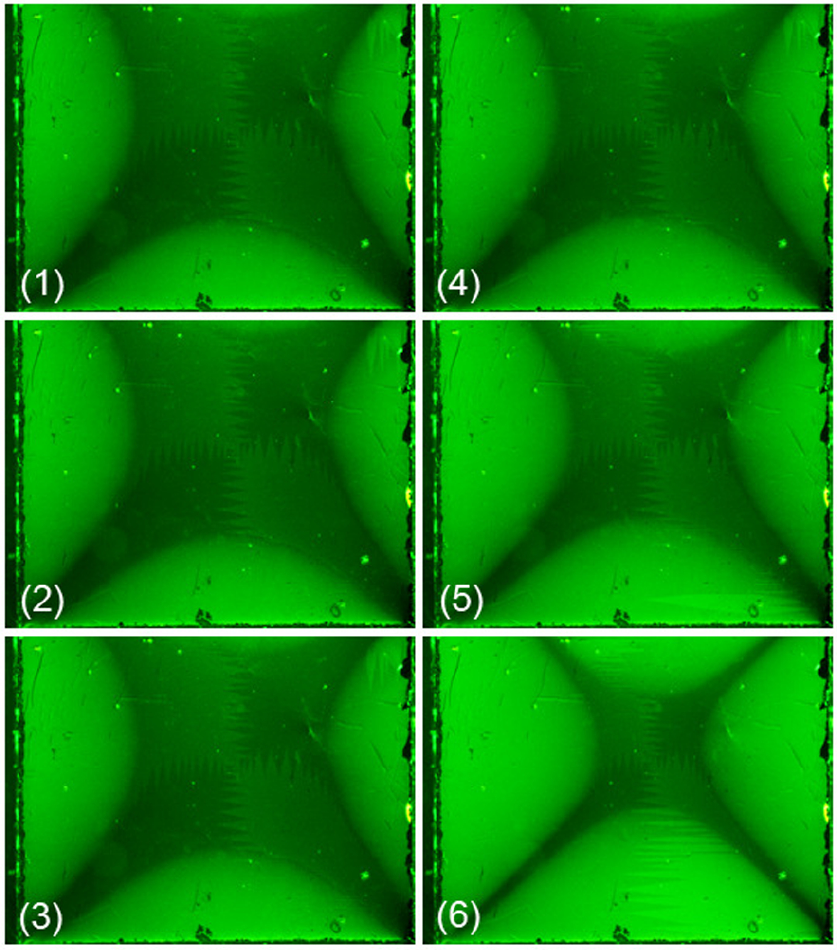}%
\caption{(Color online) MOI, showing the progression of the flux front as the microwave field power is increased. The images correspond to the microwave powers denoted by the arrows in Fig.~\protect\ref{fig5}(a): 1. No microwave field; 2. 10 dBm; 3. 14 dBm; 4. 17 dBm; 5. 20 dBm; 6. 24 dBm.}
\label{fig9}%
\end{figure}

 \begin{figure}[b]%
\includegraphics[width=1.0\linewidth]{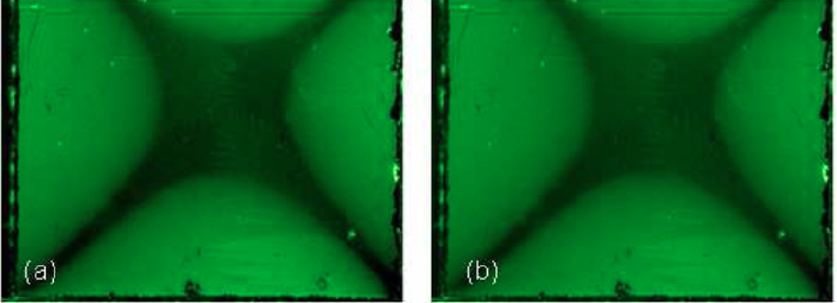}%
 
\caption{(Color online) Identical flux magnetic flux penetration into the YBa$_{2}$Cu$_{3}$O$_{7-\delta}$ thin film exposed to a dc field of 130 Oe (a) at the sample holder temperature $T = 60$ K, with applied microwave power $P_{rf}$ = 17 dBm; (b) at the higher temperature of the sample holder $T = 62.586 $ K, chosen such that $\Delta f = 33.245$ MHz as in (a), but with  $P_{rf}$ = 10 dBm. }
\label{fig9}%
\end{figure}

\subsection{MOI with microwave and dc magnetic field}

A second series of experiments was performed by adding a constant magnetic field to the rf  field. The superconducting sample was again zero-field cooled to 60 K, following which a dc magnetic field of 130 Oe was applied perpendicularly to the film surface. This leads to the magnetic flux penetration illustrated in Fig.~\ref{fig7}.  Next, the microwave input was switched on in order to excite the dielectric resonator. The nominal temperature of the cold head and the dc magnetic field remain constant during the experiment. The microwave power was gradually increased in steps of approximately 3 dBm. At each input microwave power level the $Q$-factor and the resonant frequency of the resonator was measured and the corresponding MO image recorded. The obtained results concerning the power dependence of the surface resistance and the resonant frequency are again plotted in Figure~\ref{fig5}. We observe the usual power dependence of the surface resistance of YBa$_{2}$Cu$_{3}$O$_{7-\delta}$ thin films, {\em i.e.}, a nonlinear increase of $R_{s}$ as function of $P_{rf}$. As shown by temperature dependence of $f_{0}$ depicted in  Fig.~\ref{fig8}, the simultaneous increase of the resonant frequency of the dielectric, here, by up to $\delta f  = 33.245$ MHz for the highest injected power, can be understood as the result of local heating of the YBa$_{2}$Cu$_{3}$O$_{7-\delta}$ film.\cite{Kermorvant2009}   

The simultaneously acquired MO images show no progression of the flux front in the linear regime of $R_{s}( P_{rf})$, but pronounced enhanced dc magnetic flux penetration in the nonlinear regime. A mechanism for ac magnetic field--induced dc flux penetration into thin superconducting films, the so-called vortex lattice shaking,  was presented in Refs.~\onlinecite{Brandt2002,Mikitik2003,Mikitik2004,Mikitik2005}. The conditions for  vortex shaking to be effective are that the ac field strength be sufficient to drive the film into the critical state, which is the case here, and that the rf screening currents be of opposite polarity on the top and bottom film surfaces.  In the present experimental configuration, the  latter condition is not satisfied: the use of the $TE_{011}$ cavity mode means that the rf magnetic field to which the film is subjected induces screening currents of the same polarity on the top and bottom film surface, opposite in direction to the electric field in the rutile cavity.

We therefore surmise that the progression of the flux front is due to the decrease of the critical current density $j_{c}$ associated with the increase in  temperature of the YBa$_{2}$Cu$_{3}$O$_{7-\delta}$ film, a conjecture that is checked by increasing the nominal temperature of the cold head and the sample holder using the incorporated heater, all the while maintaining a low  microwave input power of 10 dBm. The temperature is adjusted so as to precisely yield a frequency change of the resonator $\delta f  = 33.245$ MHz. We find that this corresponds to a temperature increase of $\Delta T = 2.586$~K. The MOI (Fig. ~\ref{fig9}) shows that the flux penetration precisely corresponds to that previously obtained by increasing the microwave power, showing that local heating is indeed at the origin of the enhanced flux penetration and the nonlinear dependence $R_{s}$ ( $P_{rf}$ ).

\section{Summary and Conclusions}

We have developed a experimental set-up that allows for the simultaneous imaging of magnetic flux penetration into superconducting samples using the magneto-optical technique, and the measurement of their surface resistance in the range 1 -- 10 GHz. The method was applied to YBa$_{2}$Cu$_{3}$O$_{7-\delta}$ thin films. No signature of the microwave magnetic field could be observed in magneto-optics. However, the application of a high-power microwave magnetic field significantly enhances dc flux penetration due to local heating of the superconducting film. 

\begin{acknowledgments}

This work was partially funded by the French National Research Agency ANR, under contract number ANR-07-BLAN-0242 ÔÔSURFÕÕ.

\end{acknowledgments}

\end{document}